\documentclass[aps,twocolumn,prl]{revtex4}

\pdfoutput=1

\usepackage[utf8]{inputenc}
\usepackage{lmodern}
\usepackage{hyperref}
\usepackage{amsmath}
\usepackage{amssymb}
\usepackage{graphicx}
\usepackage{xcolor}
\usepackage{color}
\usepackage{amssymb}
\usepackage{amsthm}
\usepackage{epstopdf} 
\usepackage{natbib}
\usepackage{dcolumn}
\usepackage{amsfonts}
\usepackage{bm}

\newcommand{\wcm}{\,\mathrm{W/cm}^2}
\newcommand{\jcm}{\,\mathrm{J/cm}^2}

\newcommand{\ps}{\,\mathrm{ps}}

\newcommand{\micron}{\,\mu\mathrm{m}}

\begin{document}

\title{Laser wakefield acceleration driven by few-cycle laser pulses in overdense plasmas}

\author{N. Zaïm,$^{1,}$}
\email{neil.zaim@ensta-paristech.fr}
\author{F. Böhle$^1$, M. Bocoum$^1$, A. Vernier$^1$, S. Haessler$^1$, X. Davoine$^2$, L. Videau$^2$, J. Faure$^1$ and R. Lopez-Martens$^1$}
\address{$^1$LOA, ENSTA ParisTech, CNRS, Ecole polytechnique, Universit\'e Paris-Saclay, Palaiseau, France}
\affiliation{$^2$CEA, DAM, DIF, F-91297 Arpajon, France }

\begin{abstract}

We measure the emission of energetic electrons from the interaction between ultrashort laser pulses and a solid density plasma in the relativistic regime. We detect an electron beam that only appears with few-cycle pulses ($<10$ fs) and large plasma scale lengths (L $>\lambda_0$). Numerical simulations, in agreement with the experiments, reveal that these electrons are accelerated by a laser wakefield. Plasma waves are indeed resonantly excited by the few-cycle laser pulses in the near-critical-density region of the plasma. Electrons are then injected by ionization into the plasma waves and accelerated to relativistic energies. This study provides an unprecedented insight into the physics happening in the few-cycle regime.

\end{abstract}

\maketitle

Since the recent advent of ultrahigh intensity lasers, bright particle and radiation sources with femtosecond duration have been developed from relativistic laser-plasma interactions. These new sources are expected to find applications in various fields including medicine, imaging and ultrafast probing of matter~\cite{malk08,albe16}. Laser wakefield acceleration (LWA) is an efficient process for driving relativistic electron beams with few femtosecond durations~\cite{lund11} and 100~MeV to multi-GeV energies~\cite{faur04,wang13,leem14}. In this scheme, the laser pulse ponderomotive force drives a high amplitude plasma wave that is able to trap and accelerate electrons over very short distances~\cite{esar09}. This process usually takes place in mm-scale underdense plasmas but is quite inefficient in overdense plasmas. Indeed, for solid-density plasmas, the processes responsible for transferring the laser energy to particles and radiation are radically different. Understanding the pathways and mechanisms of energy transfer to the plasma electrons is a complex and fundamental question that has implications for ion acceleration~\cite{macchi13} and high harmonic generation~\cite{thau10}. 

It is well known that the plasma density profile at the front surface is a key parameter that can dramatically transform the nature of the interaction~\cite{kaha13}. When the plasma scale length is short compared to the laser wavelength ($L<\lambda_0/10$), the physics is now fairly well understood. At moderate intensity, vacuum heating~\cite{brun87} is the dominant mechanism for energy transfer to the electrons. At relativistic intensities, the physics becomes more complex: the laser field triggers a periodic push-pull motion of the front surface that follows the sign of the laser field~\cite{gono11,thev16}. This nonlinear periodic motion leads to high harmonic generation via the relativistic oscillating mirror mechanism~\cite{lichters96,Baeva2006} and also results in electron ejection from the front surface~\cite{thev15,thev16,ceaprx}. Ejected electrons are subsequently injected into the reflected laser pulse and can gain large amounts of energy directly from the reflected laser field~\cite{thev15}. In this regime, electron emission has been reported to be optimal when the gradient scale length is on the order of $\sim \lambda_0/10$, with electron energies ranging from 100 keV to multi-MeV, depending on the laser intensity~\cite{mord09,tian12,thev15,bocoum16,ceaprx}. 

For longer gradient scale lengths, there is no unified description of energy transfer and electron acceleration. There is a wide disparity of experimental results and various mechanisms have been proposed, including resonant absorption~\cite{li01,cai04}, J$\times$B heating~\cite{malk96b}, ponderomotive acceleration~\cite{zhan04}, stochastic heating~\cite{tomm04,feis17}, acceleration by surface quasistatic fields~\cite{li06} or direct laser acceleration~\cite{ma17}. However, it is still unclear what mechanisms actually arise in experiments and the precise experimental conditions under which they appear is not known. This may be due to the lack of control and measurement of the density gradients, which makes the interpretations difficult. In this letter, we show that by precisely controlling the density gradient and using few-cycle laser pulses with durations as short as 3.5~fs, we clearly identify for the first time a regime where LWA occurs in a near-critical density plasma. We find that this regime, which results in the emission of a stable low-divergence electron beam, only occurs for few-cycle laser pulses, clearly showing that extremely short pulse durations provide access to new acceleration regimes.

The experiments are performed with the Salle Noire laser system at the Laboratoire d'Optique Appliquée (LOA). The laser delivers 2.6-mJ pulses at 1-kHz repetition rate with an extremely high temporal contrast $> 10^{10}$~\cite{jull14}. The 800~nm, 24~fs laser pulses are post-compressed in a helium-filled stretched hollow-core fiber~\cite{boeh14,ouil18}. The pulse duration can be tuned by changing the pressure in the fiber, thereby providing near Fourier transform limited pulses from 3.5~fs to 24~fs. The laser beam is focused down to a 1.75$\micron$ FWHM spot resulting in peak intensities ranging from $2.3 \times 10^{18} \wcm$ ($a_0 \simeq 1$) for 24 fs pulses to $1.6 \times 10^{19} \wcm$ ($a_0 \simeq 2.7$) for 3.5~fs pulses. Here, $a_0$ is defined as the normalized amplitude of the peak laser field: $a_0 = E_\mathrm{MAX}/E_0$ with $E_0 = m_\mathrm{e} c \omega_0 / e$ where $\omega_0$ is the laser frequency, $c$ is the speed of light in vacuum and $m_\mathrm{e}$ and $e$ are the electron mass and charge respectively. 

In this experiment, p-polarized pulses impinge on an optically flat fused silica target with an incidence angle $\theta_i = 55^\circ$. A spatially overlapped prepulse, created by picking off $\approx 4 \%$ of the main pulse through a holey mirror, is focused to a much larger 13$\micron$ FWHM spot in order to generate a transversely homogeneous plasma that expands into vacuum. The plasma density profile during the interaction is controlled by varying the delay, $\Delta t$, between the prepulse and the main pulse. The density scale length is estimated experimentally using spatial domain interferometry and assuming isothermal expansion~\cite{bocoum15}. Backward electron emission is measured using a Lanex screen, protected by a 13$\micron$ thick Al-foil, which detects electrons with energies $> 150\,$keV. The Lanex screen was calibrated prior to the experiment using a 3-MeV RF accelerator. The absolute charge is estimated from the electron energy spectrum obtained from the PIC simulations described below combined with the known spectral response of the Lanex screen. The resulting uncertainty is on the order of $50\%$. The angular electron distribution in the backward direction is recorded for $-3^\circ < \theta < 75^\circ$ and $-15^\circ < \phi < 15^\circ$ where $\theta$ and $\phi$ are the angles with respect to target normal respectively in the incidence and transverse planes.

\begin{figure}
\includegraphics[width=1.\columnwidth]{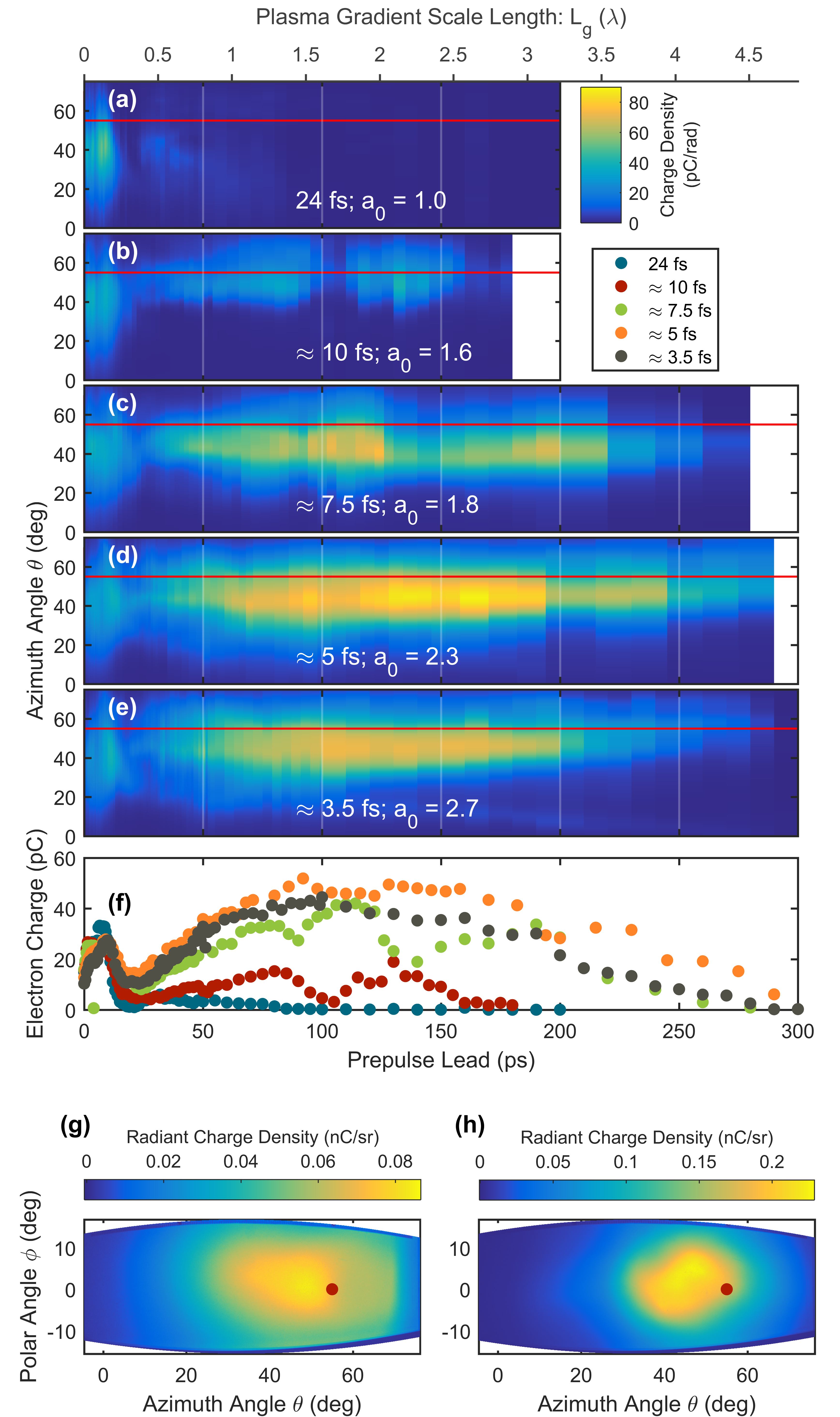}
\caption{(a)-(e) Electron angular distribution integrated over the $\phi$ angle as a function of the delay between the prepulse and the main pulse for respective pulse durations of 24, 10, 7.5, 5 and 3.5 fs. (f) Total ejected charge as a function of the delay between prepulse and main pulse. (g), (h) Typical electron angular distribution respectively in the short ($\Delta t = 9 \ps$) and long ($\Delta t = 140 \ps$) plasma scale length regimes. The red lines and dots mark the specular direction.}
\label{FigExperiment}
\end{figure}

Figures~\ref{FigExperiment}(a)-\ref{FigExperiment}(f) show the measured electron signal as a function of the delay between the prepulse and the main pulse for 5 different laser pulse durations. We first find a strong electron emission for short delays ($\Delta t < 20$ ps), corresponding to a sharp plasma-vacuum interface. This emission, detected for every pulse duration, is optimal for a delay $\Delta t \approx 9$ ps, i.e.  $L<\lambda_0/5$. In this physical regime, the push-pull mechanism mentioned in the introduction is responsible for the ejection of electrons from the plasma~\cite{thev15,thev16,bocoum16}. A typical electron angular distribution in this case is displayed in Fig.~\ref{FigExperiment}(g), showing a broad divergence angle of $\approx 50^\circ$.

As the delay is further increased, the detected charge drops (10 ps $< \Delta t < 30$ ps), and then rises again for longer delays ($\Delta t > 50$ ps). This time however, electrons are only emitted when few-cycle pulses ($\leq$10~fs) are used. Note that chirping a few-cycle pulse to increase its duration results in a similar decline of the electron signal~\cite{supp}. This is thus a very distinct physical regime, in which the gradient length is much larger ($L>\lambda_0$) and the duration of the laser pulse plays a major role. In this case, the obtained electron beams have more charge and a narrower divergence angle of $\approx 25^\circ$ as is visible in Fig.~\ref{FigExperiment}(h). The electrons are emitted near the specular direction, with a slight shift towards the normal direction. The detected signal is very stable over a wide range of delays (50 ps$< \Delta t <$200 ps), indicating that the electron ejection mechanism is not highly sensitive to the exact shape of the plasma density profile. This mechanism also appears to be strongly nonlinear, as the electron signal at long delays drops much faster than the signal at short delays when reducing the laser intensity~\cite{supp}. 

\begin{figure}
\includegraphics[width=1.\columnwidth]{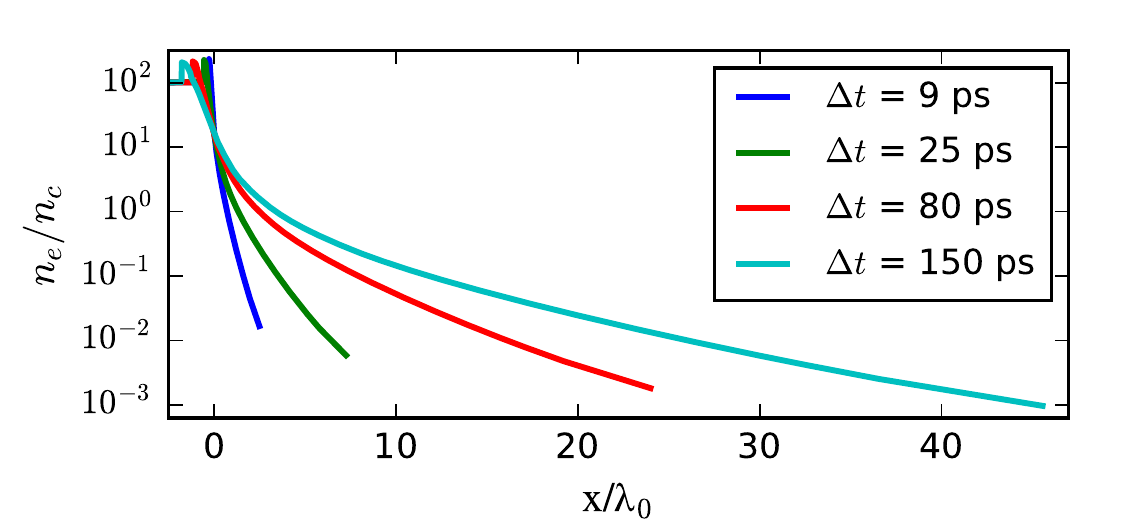} 
\caption{Results from 1D hydrodynamical simulations. Normalized electron density $n_\mathrm{e} / n_\mathrm{c}$ as a function of position $x$ for different delays after a prepulse with fluence $50 \jcm$ ionizes a solid fused-silica target. $x=0$ is the initial solid-vacuum interface position. $n_\mathrm{c} = m_\mathrm{e} \epsilon_0 \omega_0^2 / e^2$ is the critical density above which the laser cannot propagate. $\epsilon_0$ is the vacuum permittivity.}
\label{FigEsther}
\end{figure}

To understand the origin of this new electron emission process, we turn to 2D Particle-In-Cell (PIC) simulations using the code WARP~\cite{warp05}. We use the same laser parameters as in the experiments (more details are given in the Supplemental Material~\cite{supp}). A moving window is started after the interaction in order to follow the accelerated electrons far from the plasma. We took great care in providing a realistic description of the plasma density gradient. First, the plasma is initially partially ionized (up to Si$^{4+}$ and O$^{2+}$) in order to model ionization by the prepulse. The initial ionization states are estimated from the prepulse peak intensity ($\sim 10^{15} \wcm$) and the intensity thresholds for barrier-suppression ionization~\cite{gibb04} in silicon and oxygen. Further ionization by the main pulse is also taken into account in the simulations. Second, the plasma density profile is obtained by performing hydrodynamical 1D simulations with the code ESTHER~\cite{colo06}. Figure~\ref{FigEsther} shows the resulting profiles for 4 different values of the delay between the prepulse and the main pulse. Note that the density profiles are not always exponential in Fig.~\ref{FigEsther}, contrary to results from models assuming isothermal expansion. The gradient appears to have an exponential shape only for short delays (i.e. for sharp plasma-vacuum interfaces) but not for longer delays. The isothermal hypothesis likely fails due to radiation and convection losses on these longer timescales. In our case, the electron beam appears for long delays $\Delta t$ and we therefore use the density profiles shown in Fig.~\ref{FigEsther} as inputs for the PIC simulations.

\begin{figure*}[t]
\includegraphics[width=2.\columnwidth]{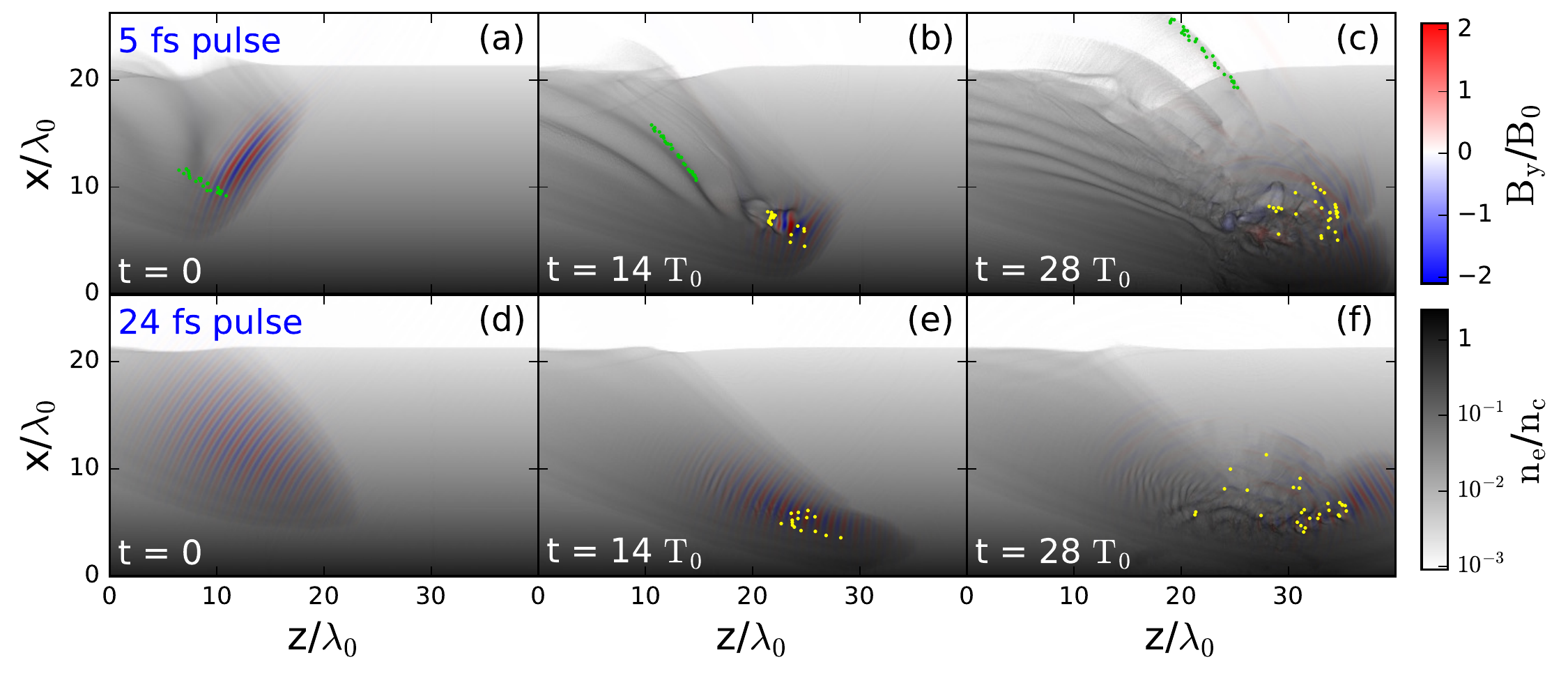}
\caption{Laser magnetic field and electron density from PIC simulations with a large plasma scale length ($\Delta t$ = 80 ps) and a pulse duration of (a)-(c) 5 fs or (d)-(f) 24 fs. The green and yellow dots show a sample of ejected electrons. $T_0$ is the laser optical oscillation period and $B_0=E_0/c$ with $E_0$ defined earlier. See also Supplemental Material movies for a more comprehensive view of the simulations.~\cite{supp}}
\label{FigSnapshots}
\end{figure*}

Snapshots from two different PIC simulations are shown in Fig.~\ref{FigSnapshots}. Both simulations use the plasma density profile obtained with a delay $\Delta t$ = 80 ps (i.e. the red curve in Fig.~\ref{FigEsther}), a value for which the electron beam is detected in the experiments. The pulse duration is either 5 fs or 24 fs, resulting in peak intensities of $1 \times 10^{19} \wcm$ ($a_0 = 2.15$) and $2.1 \times 10^{18} \wcm$ ($a_0 = 0.98$) respectively.

The first striking feature is the formation of high amplitude plasma waves in the wake of the 5-fs pulse. Their wavefront is bent by the density gradient, akin to the plasma waves generated by Brunel electrons in the coherent wake emission mechanism of high-harmonic generation~\cite{thau10}. Even though these wakefields appear in the whole region where the 5-fs pulse propagates, inside which the density ranges from $n_\mathrm{c}$/1000 to $n_\mathrm{c} \cos^2 \theta_i \sim 0.3 n_\mathrm{c}$~\cite{krue88}, they are completely absent in the 24-fs pulse simulation. This can be easily explained by the fact that wakefield excitation is optimal at the resonance condition, i.e. when the pulse duration is on the order of half the plasma wavelength: $\tau\simeq \lambda_p/2c$~\cite{esar09}. This gives a resonant density of $n_c/14$ for 5-fs pulses, versus $n_c/300$ for 24-fs pulses, explaining why large wakefields appear for the few-cycle pulse only (see also Supplemental Material~\cite{supp}). 

\begin{figure}
\includegraphics[width=1.\columnwidth]{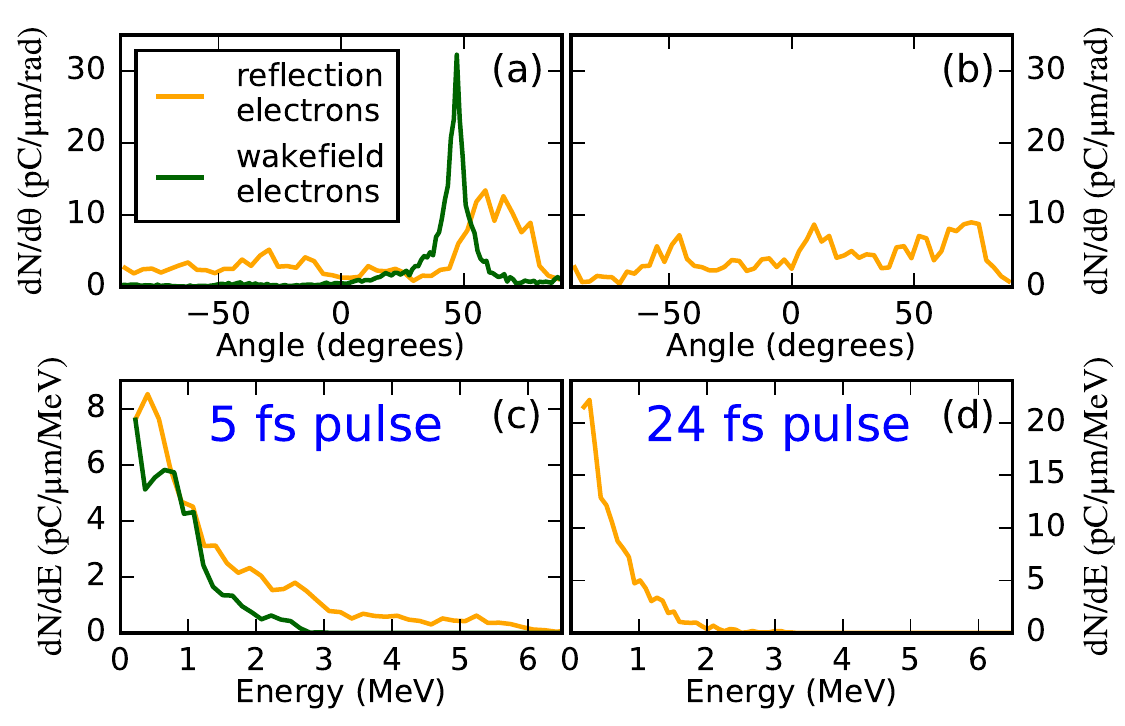}
\caption{(a),(b) Angular and (c),(d) energy distribution of the two families of electrons that are ejected in the (a),(c) 5 fs and (b),(d) 24 fs simulation. The distributions are obtained at the end of the simulation, long after the interaction.}
\label{FigDistributions}
\end{figure}

Some electrons, represented in green in Fig.~\ref{FigSnapshots}, are trapped and accelerated by the plasma waves in the PIC simulations. The angular and energy distribution of these LWA electrons is shown in the green curves of Fig.~\ref{FigDistributions}(a) and Fig.~\ref{FigDistributions}(c) respectively. Their energy spectrum extends to $\approx$ 2.5 MeV and their total ejected charge is $\approx$ 7 pC/$\micron$. These electrons have the same angular distribution as the electrons detected at long delays in experiments (see Fig.~\ref{FigExperiment}(d) for $\Delta t$ = 80 ps). Moreover, as in experiments, these electrons only appear for few-cycle pulses. We therefore conclude that the electron beam detected at long delays in experiments originates from LWA. Even though this mechanism was previously suggested for electron acceleration in overdense plasmas~\cite{cai04,mao15}, the evidence supporting these claims remained inconclusive. To our knowledge, this is the first time that LWA from solid density plasmas is clearly identified in experiments.

In the simulations, the LWA electrons originate from the L-shell of silicon. They have high binding energies (from $\approx$ 150 eV to $\approx$ 500 eV) and can therefore only be ionized by the huge electric fields inside the main laser pulse. The fact that only electrons ionized in the center of the pulse are accelerated suggests that injection by ionization, a well-known mechanism in underdense plasmas~\cite{mcgu10,pak10}, is responsible for trapping the electrons into the wakefields. Taking field ionization into account is therefore needed to properly describe the injection of electrons into the plasma waves and more generally to correctly model laser interactions with overdense plasmas when the plasma scale length is large.

Another family of electrons, shown in yellow in Fig.~\ref{FigSnapshots} and labelled ``reflection electrons'', is ejected from the plasma in the simulations. These electrons are accelerated at the reflection point of the laser, where the density is $n_\mathrm{c} \cos^2 \theta_\mathrm{i}$. Their angular and energy distributions are displayed in the yellow curves of Fig.~\ref{FigDistributions}. This family of electrons, which appears for both 5-fs and 24-fs pulses and has a very large angular divergence spreading across all directions, is not detected in experiments. This fact shows that our 2D PIC simulations do not accurately reproduce the ratio between the two populations. This second family of electrons would most likely be attenuated in more accurate but more costly 3D simulations. However, this is not a major concern because the simulations explain the main experimental observations, i.e. a well-defined beam of LWA electrons that appears only for extremely short pulse duration.

It is also worth noting that the same qualitative results are found when the simulations are performed with an exponential density profile with $L = 3\lambda_0$, thus confirming our previous observation that the electron ejection mechanism is not highly sensitive to the exact shape of the plasma density profile. Another interesting point is that a similar trend can be found when the intensities are interchanged in the simulations (i.e. when the 5-fs simulation is carried out with a peak intensity of $2.1 \times 10^{18} \wcm$, while the 24-fs simulation is performed with a $1 \times 10^{19} \wcm$ peak intensity). In this case, even though the laser pulse energy is 25 times lower in the 5-fs simulation, a small amount of electrons remain laser wakefield accelerated while there is still no plasma wave formation in the 24-fs simulation~\cite{supp}. This clearly shows that the generation of the electron beam is due to the decrease in pulse duration rather than to the associated increase in intensity. 

In conclusion, we detect an electron beam that only appears for few-cycle pulses and large plasma scale lengths. Particle-In-Cell simulations successfully explain the experimental results: the detected electrons are injected by ionization into wakefields formed behind the pulse. These plasma waves can only be efficiently excited by few-cycle pulses at these near-critical densities, explaining why this new electron emission mechanism is only observed with extremely short pulses. This work offers a better understanding of the interaction between ultraintense laser pulses and overdense plasmas and confirms that new physical phenomena happen in the few-cycle regime, most of which are still to be discovered and understood.

This work was funded by the European Research Council under Contract No. 306708 (ERC Starting Grant No. FEMTOELEC), the Région Île-de-France (under Contract No. SESAME-2012- ATTOLITE), the Agence Nationale pour la Recherche (under Contracts No. ANR-11-EQPX-005-ATTOLAB and No. ANR-14-CE32-0011-03), the Extreme Light Infrastructure-Hungary Non-Profit Ltd (under Contract No. NLO3.6LOA) and the LabEx PALM (ANR-10-LABX-0039). This work was granted access to the HPC resources of CINES under the allocation A0020506057 made by GENCI.

\bibliographystyle{apsrev}

\end{document}


\title{Supplemental material for: Laser wakefield acceleration driven by few-cycle laser pulses in overdense plasmas}

\author{N. Zaïm,$^{1,}$}
\email{neil.zaim@ensta-paristech.fr}
\author{F. Böhle$^1$, M. Bocoum$^1$, A. Vernier$^1$, S. Haessler$^1$, X. Davoine$^2$, L. Videau$^2$, J. Faure$^1$ and R. Lopez-Martens$^1$}
\address{$^1$LOA, ENSTA ParisTech, CNRS, Ecole polytechnique, Universit\'e Paris-Saclay, Palaiseau, France}
\affiliation{$^2$CEA, DAM, DIF, F-91297 Arpajon, France }

\maketitle

\section{Experiments with reduced intensity}

Figure~\ref{ExpAperture} shows the experimental results obtained when aperturing the laser beam, which resulted in an intensity reduced to $11\%$ of its original value. The signal at long gradient (originating from LWA) drops much faster than the signal at short gradient (coming from the push-pull mechanism) when the intensity is decreased. This may be due to the nonlinear intensity dependence of the field-ionization responsible for the injection of electrons into the plasma waves.

\renewcommand{\thefigure}{S\arabic{figure}}

\begin{figure}
\makebox[\textwidth][c]{\includegraphics[width=0.85\columnwidth]{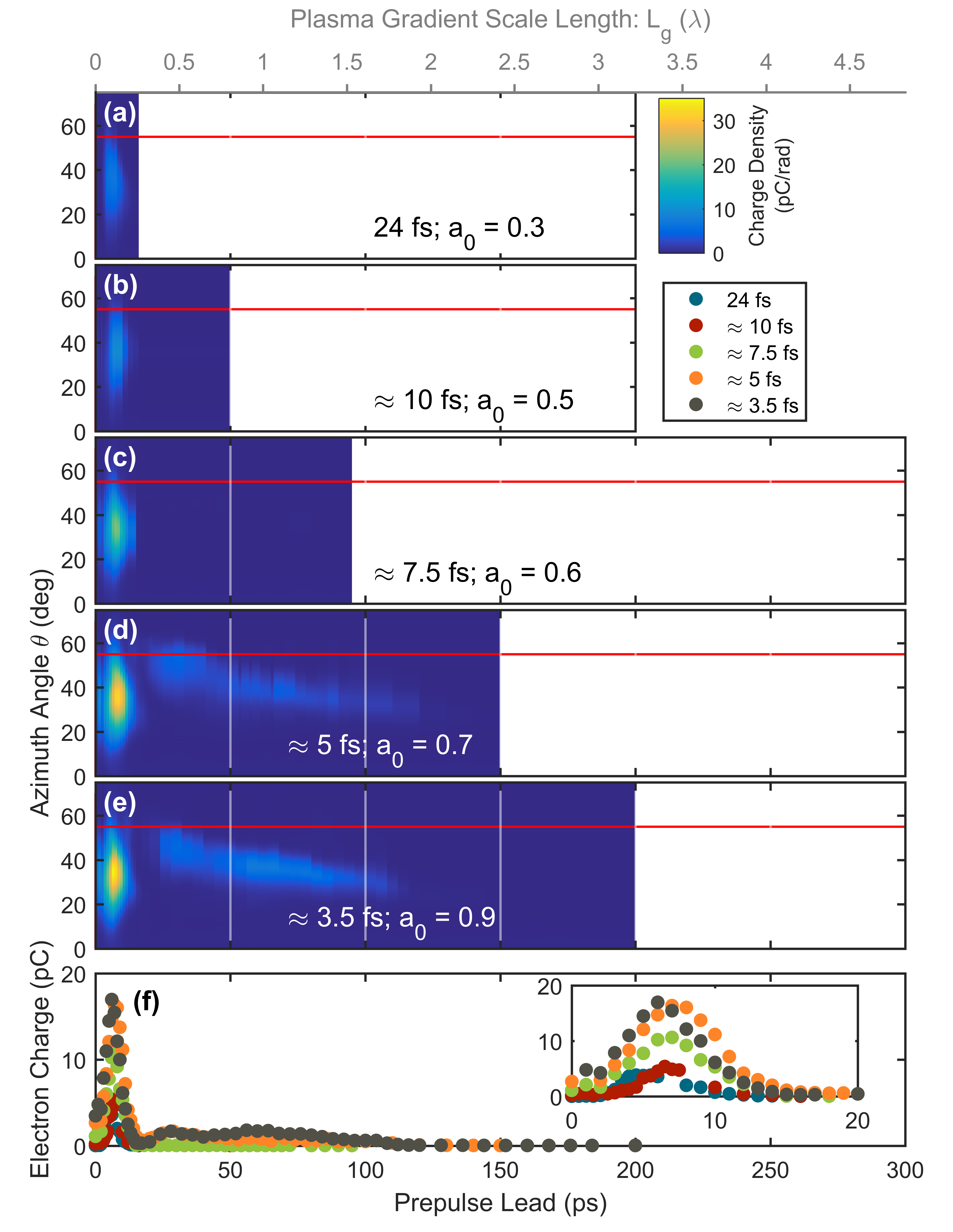}}
\caption{(a)-(e) Electron angular distribution integrated over the $\phi$ angle as a function of the delay between the prepulse and the main pulse for respective pulse durations of 24, 10, 7.5, 5 and 3.5 fs and reduced intensity. (f) Total ejected charge as a function of the delay between prepulse and main pulse.}
\label{ExpAperture}
\end{figure}

\section{Experiments with chirped pulses}

\begin{figure}
\makebox[\textwidth][c]{\includegraphics[width=0.85\columnwidth]{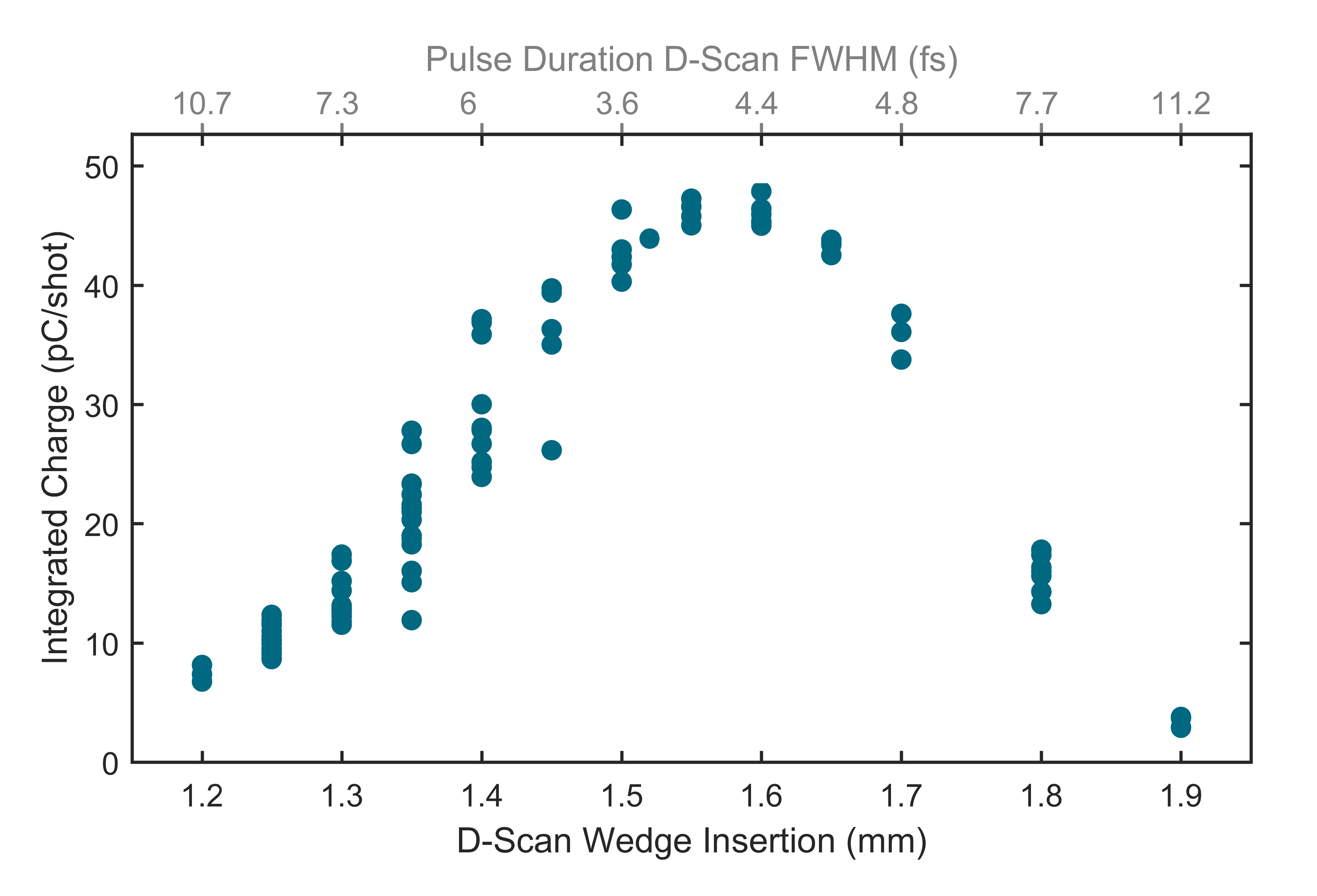}}
\caption{Total detected charge as a function of the pulse duration for a delay $\Delta t = 90 \ps$ between the prepulse and the main pulse.}
\label{ExpChirp}
\end{figure}

The experimental integrated electron signal in the long gradient regime ($\Delta t = 90 \ps$) is shown in Fig.~\ref{ExpChirp} as a function of the pulse duration. Here, the pulse duration is tuned from 3.5 fs up to 11.2 fs by chirping both positively and negatively the driving laser pulse. The results are fully consistent with the ones presented in the main text: the electron beam in the long gradient regime is only detected when extremely short laser pulses ($< 10 \fs$) are used.

\section{Modeling 2D few-cycle laser pulses}

The laser field used as an input in the PIC simulations is obtained from the paraxial 2D gaussian beam and assuming a Poisson-like spectrum, used to describe few-cycle pulses\cite{caro99,apri10}. The expression of the non-zero components of the laser field is then the following:

\begin{equation}
\label{ExInt}
E_x(\mathbf{r},t) = E_0 \int F(\omega) \sqrt{\frac{w_0}{w(z)}} \exp \left(-\left(\frac{x}{w(z)}\right)^2\right) \cos \left( \omega t - kz + \frac{1}{2} \arctan \left( \frac{z}{z_R}\right) -\frac{kx^2}{2R(z)} \right) d\omega
\end{equation}

\begin{equation}
B_y(\mathbf{r},t) = \frac{E_x(\mathbf{r},t)}{c} 
\end{equation}

z is the direction of propagation of the laser, x is the direction of polarization, $E_0$ is the amplitude of the field, c is the speed of light in vacuum, $\omega$ is the laser frequency, $k=\omega/c$ is the laser wavenumber, $w_0$ is the beam waist radius, $z_R=\frac{kw_0^2}{2}$ is the Rayleigh length, $w(z) = w_0 \sqrt{1+(\frac{z}{z_R})^2}$ is the beam radius, $R(z) = z+\frac{z_R^2}{z}$ is the radius of curvature and $F(\omega)$ is the Poisson-like pulse spectrum $F(\omega)$, given by:

\begin{equation}
F(\omega) = \mathrm{e}^{\mathrm{i}\phi_0} \left( \frac{s}{\omega_0}\right)^{s+1} \frac{\omega^s\mathrm{e}^{-s\frac{\omega}{\omega_0}}}{\Gamma(s+1)} H(\omega)
\end{equation}

Where $\omega_0$ is the frequency with maximum spectral amplitude, $\phi_0$ is the absolute phase, $\Gamma$ and $H$ are respectively the gamma and the Heaviside functions and s is a parameter linked to the pulse duration in FWHM of intensity $\tau$ by the following relation: $\omega_0\tau = \sqrt{2}s [4^{1/(s+1)}-1]^{1/2}$~\cite{caro99}. The integral in equation~\ref{ExInt} is calculated in the complex space assuming that the pulse is isodiffracting, i.e. that every frequency has the same Rayleigh length $z_R$. The resulting electric field is the real part of the following expression:

\begin{equation}
\widetilde{E_x}(\mathbf{r},t) = E_0 \mathrm{e}^{\mathrm{i}\phi_0} \sqrt{\frac{iz_R}{\widetilde{q}(z)}} \left( 1 - \frac{i}{s} \left( \omega_0 t -k_0z - \frac{k_0 x^2}{2 \widetilde{q}(z)} \right) \right)^{-(s+1)}
\end{equation}

Where $\widetilde{q}(z) = z + iz_R$ and $k_0 = \omega_0 / c$. Note that this expression is only used to inject the laser pulse at the edge of the simulation box. Higher order terms, such as the longitudinal electric field, will naturally appear in the simulations as the laser propagates on the grid, even though they are not included in this expression.

\section{Simulations parameters}

This section gives details about the parameters used in the simulations presented in the main text.

The numerical parameters are the following: space steps $\Delta x = \Delta z = \lambda_0 / 71$ where $\lambda_0$ = 800 nm, timestep $\Delta t = T_0/101$ where $T_0=\lambda_0/c$, box size $N_x\times N_z = 4376 \times 9579$, 4 particles per cell per species (electrons, Si$^{4+}$ ions and O$^{2+}$ ions) initially. After further ionization by the main pulse, the number of electron per cell ranges from 4 to 60, depending on the final degree of ionization. Increasing the resolution or the number of particles per cell does not change the simulation results.

The laser parameters are the following: $\omega_0 = 2\pi/T_0$, $z_R = 8,84 \micron$ (corresponding to $w_0 = 1,5 \micron$), $s=50$ or $s=1154$ (corresponding to respective pulse durations at FWHM of intensity of $\tau = 5$ fs and $\tau = 24$ fs), $a_0 = 2.15$ in the 5 fs simulation and $a_0 = 0.98$ in the 24 fs simulation. The laser focus point is located at the position of the initial solid-vacuum interface. The laser is thus reflected slightly before focus because of the plasma expansion. 

\section{Simulations with intensities interchanged}

\begin{figure}
\makebox[\textwidth][c]{\includegraphics[width=1.2\columnwidth]{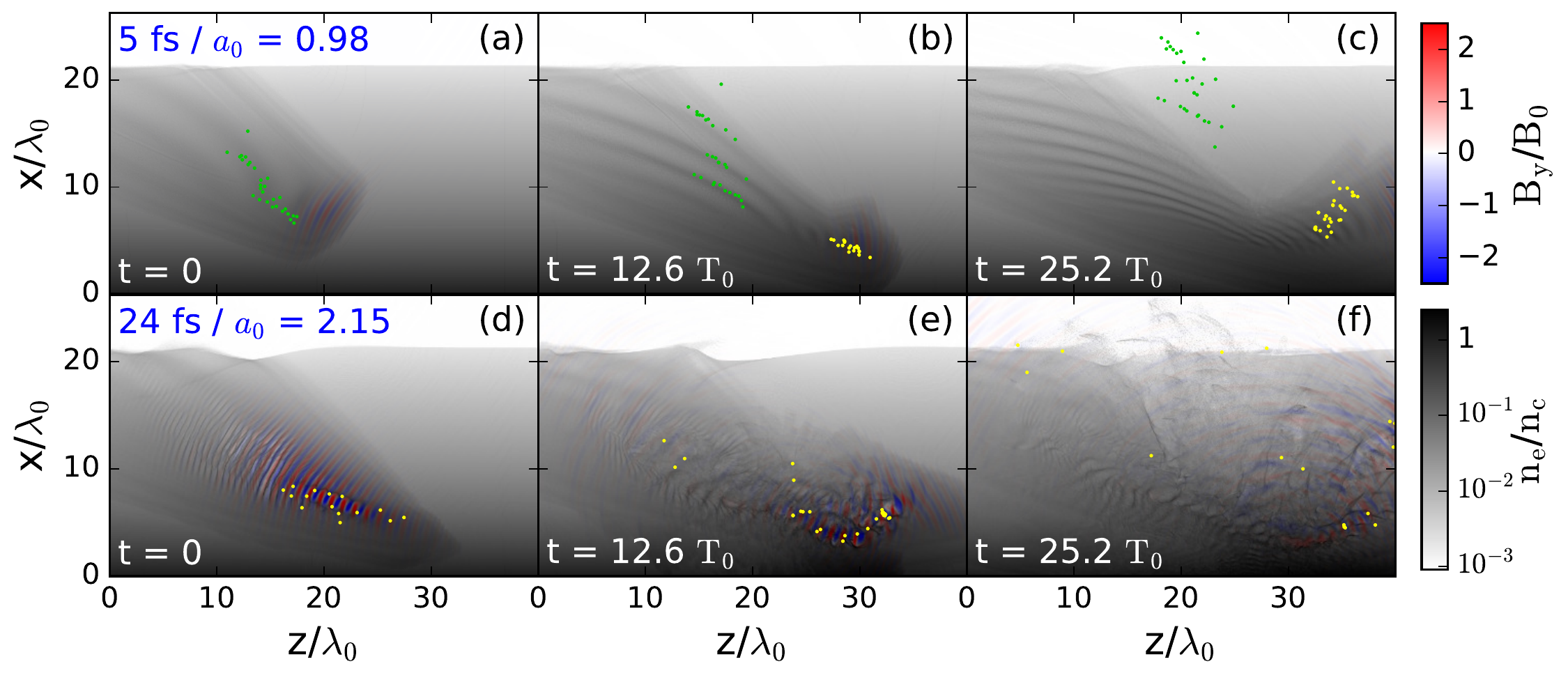}}
\caption{Laser magnetic field and electron density from PIC simulations with a large plasma scale length ($\Delta t = 80$ ps), a pulse duration of (a)-(c) 5 fs or (d)-(f) 24 fs and a normalized field amplitude $a_0$ of (a)-(c) 0.98 or (d)-(f) 2.15. The green and yellow dots show a sample of ejected electrons that are originating from silicon ions.}
\label{SimInterchanged}
\end{figure}

Figure~\ref{SimInterchanged} shows the simulation results with interchanged intensities, i.e. when the 5-fs laser pulse has a peak intensity of $2.1 \times 10^{18} \wcm$ and the 24-fs laser pulse has a peak intensity of $1 \times 10^{19} \wcm$. Interestingly, the main results stay the same: a small amount ($\approx 300 \fc / \mic$) of electrons are laser wakefield accelerated in the 5-fs simulation while no plasma wave formation is observed in the 24-fs simulation. This clearly shows that the decrease in pulse duration, rather than the associated increase in intensity, is responsible for the appearance of the electron beam at long gradients.

\section{Estimation of wakefield amplitude from theory}

We can estimate the amplitude of the wakefields generated in our experiments from the 1D nonlinear theory of wakefield generation~\cite{bere92,esar93}. In this framework, the normalized wakefield electric potential $\phi$ generated by a laser pulse in a plasma with uniform density $n_\mathrm{e}$ is obtained by solving the following differential equation:

\begin{equation}
\dfrac{\partial^2 \Phi}{\partial \xi^2} = k_\mathrm{p}^2 \gamma_\mathrm{p}^2 \left( \beta_\mathrm{p} \left( 1 - \dfrac{1+\hat{a}^2/2}{\gamma_\mathrm{p}^2 ( 1 + \phi )^2} \right)^{-1/2} -1 \right)
\label{1Dnonlinear}
\end{equation}

Here, $\xi = z - v_\mathrm{g}t$, where $z$ is the direction of propagation of the laser and $v_\mathrm{g}$ is the laser group velocity in the plasma, $k_\mathrm{p} = \omega_\mathrm{p}/c$, where $\omega_\mathrm{p}$ is the plasma frequency, $\beta_\mathrm{p} = v_\mathrm{g} / c$, $\gamma_\mathrm{p} = 1 / \sqrt{1-\beta_\mathrm{p}^2}$ and $\hat{a}^2$ is the averaged normalized intensity envelope of the laser, assumed in the following to be gaussian: $\hat{a}^2 = a_0^2 \exp \left( \dfrac{4 \ln (2) \xi^2 }{c^2 \tau^2} \right)$, where $\tau$ is the pulse duration (FWHM in intensity). $\phi$ is normalized by $\phi = e V / (m_\mathrm{e} c^2)$ where $V$ is the wakefield electric potential.

\begin{figure}
\makebox[\textwidth][c]{\includegraphics[width=0.75\columnwidth]{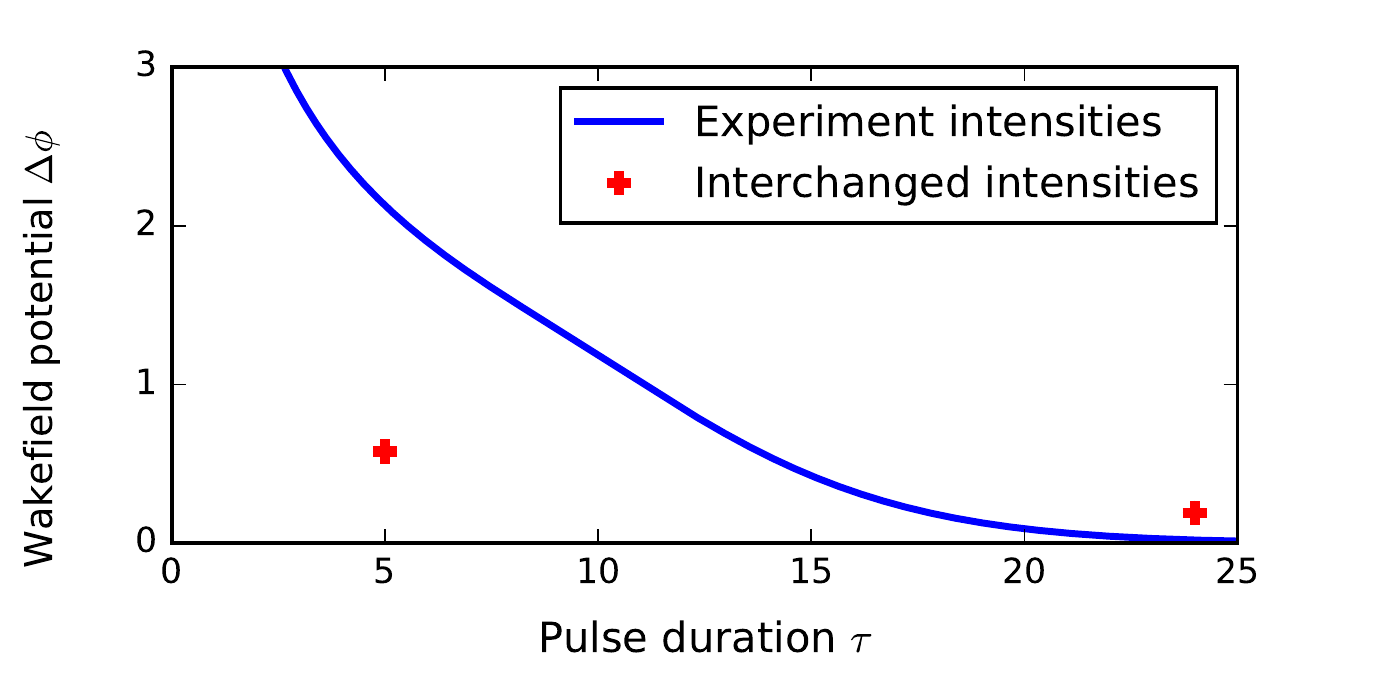}}
\caption{Normalized peak-to-peak amplitude $\Delta \phi$ of the electric potential of a laser driven wakefield in a $n_\mathrm{c}$/50 density plasma as a function of the FWHM pulse duration $\tau$ obtained using the 1D nonlinear theory of wakefield generation and (blue curve) our laser parameters or (red markers) the laser parameters used in the simulations with interchanged intensities presented in Fig.~\ref{SimInterchanged}.} 
\label{AmpWakefield}
\end{figure}

We solve equation~\ref{1Dnonlinear} numerically using our laser parameters and assuming a plasma density of $n_\mathrm{e} = n_\mathrm{c} / 50$, a typical value for which plasma waves are formed in our simulations. The resulting peak-to-peak amplitude of the generated wakefields is shown in the blue curve of Fig.~\ref{AmpWakefield} as a function of pulse duration. Unsurprisingly, only ultrashort laser pulses ($\tau < 10 \fs$) are able to excite high amplitude plasma waves. In particular, the wakefield potential amplitude is $\approx$ 100 times higher with 5-fs pulses than with 24-fs pulses. Note that the laser energy is kept constant in these calculations; the laser electric field amplitude thus scales as $a_0 \propto 1 / \sqrt{\tau}$.

The red markers in Fig.~\ref{AmpWakefield} show the wakefield amplitudes obtained when using the laser parameters used in the simulations with interchanged intensities presented in Fig.~\ref{SimInterchanged} (5 fs, $a_0 = 0.98$ and 24 fs, $a_0 = 2.15$). Altough the laser pulse energy is 25 times higher in the 24-fs pulse, the generated wakefield amplitude is 3 times higher with the 5-fs pulse. This result clearly highlights the necessity of using few-cycle pulses in order to generate plasma waves in near-critical density plasmas.

\bibliographystyle{apsrev}